\documentclass[12pt]{iopart}
\usepackage[dvips]{graphics}

\begin{document}

\title {Squeezed vacuum reservoir effect for  entanglement decay in nonlinear quantum scissors system.}

\author {A. Kowalewska-Kud{\l}aszyk$^1$, W. Leo\'nski$^2$ }
\address{$^1$Nonlinear Optics Division, Department of Physics, Adam Mickiewicz
 University, Umultowska 85, 61-614 Pozna\'n, Poland}
\address{$^2$Quantum Optics and Engineering Division, Institute of Physics, University of Zielona G\'ora, Prof.~Z.~Szafrana 4a, 65-516 Zielona G\'ora, Poland}

\ead{$^1$annakow@amu.edu.pl, $^2$wleonski@proton.if.uz.zgora.pl}


\vspace{2pc}

\begin{abstract}
We discuss the coupler system of two nonlinear oscillators excited by an external coherent field prepared in a maximally entangled state (Bell-like state). We show that as a result of the coupler interaction of the system with external broadband squeezed vacuum bath, entanglement decay dynamics can be considerably affected. Besides the phenomena of sudden entanglement death and its rebirth, a shortening (or lengthening) of the total disentanglement time $\tau_D$ can be observed, depending on the squeezing parameters. Moreover, on the example of one of the reborn entanglement cases it is shown that by changing  the values of these parameters the maximal values of the negativity for the $3\otimes 2$ system discussed can be tailored.
\end{abstract}

\maketitle
\section{Introduction}\label{intro}
Modern quantum physics still takes a very lively interest in quantum entanglement. The entanglement problems are exciting not only from the cognitive point of view, but also as a source of future applications in the field of quantum information theory.
In real situations the system which is able to generate entangled states always interacts with an external environment, and this interaction leads to the losses of entanglement in the systems considered (for instance, for discussion concerning decoherence processes for a system of two qubits in a lossy cavity see \cite{M04} \textit{and the references quoted therein}). In particular, if we deal with the cases when the system interacts with a zero-temperature reservoir, one can observe asymptotic entanglement decay (if there is no interaction between the qubits which produce an entangled state) --- meaning that entanglement decays to zero for time $\rightarrow\infty$. In contradiction, interaction with a thermal reservoir is  able to  shorten significantly that time and entanglement is destroyed in finite time (such a behaviour is referred to as the \textit{sudden entanglement death} \cite{ZHHH01,D03,YE04}). It is also known that when the entangled system interacts with squeezed vacuum bath, sudden death of entanglement can be observed \cite{ILZ07}. It should be mentioned that under certain conditions it is possible to rebuild entanglement in the system considered after its sudden death (this is called the sudden entanglement birth \cite{FT06,FT08,LRLSR08}).  Moreover, such "sudden" phenomena can be observed not only for entangling optical systems but also for other models, for instance dealing with spins. Thus, quite recently Wang \textit{et al} \cite{WMLSN10} have discussed \textit{sudden spin-squeezing death} effect.

Moreover, to avoid decoherence processes several schemes were proposed. For instance, some of them use decoherence free subspaces (DFS) method. In particular, one can show that the systems interacting with squeezed environment states taken from DFS will never show sudden death phenomena. Contrary, when the interaction involves states that are taken not from DFS the system will loose entanglement in a sudden way (with possible birth events) \cite{HO08}.

In the present paper we will focus on the decay of initially maximally entangled state generated via the coherently excited nonlinear coupler (two nonlinear, Kerr-like oscillators interacting nonlinearly with each other) system. Such couplers can exhibit numerous interesting physical phenomena not only in the domain of quantum information theory but also in other fields of quantum optics (for the thorough review of the problem concerning nonlinear couplers see for instance  review paper \cite{PP00} \textit{and the references quoted therein}, whereas for problems of nonlinear quantum-optical oscillators in general see \cite{PL95}). We shall show that we can manipulate the character of the entanglement decay to some extent via the phase of squeezed reservoir in which the system is immersed.

\section{The model and its solutions in squeezed reservoir}
The system we deal with is composed of two nonlinear oscillators labelled by $a$ and $b$, and characterised by the nonlinearity constants $\chi_{a}$ and $\chi_{b}$, respectively.  The oscillators are described by the Hamiltonian expressed by the bosonic creation and annihilation operators in the following form $\hat{H}_{nl}=\frac{\chi_a}{2}(\hat{a}^\dagger)^2\hat{a}^2+\frac{\chi_b}{2}(\hat{b}^\dagger)^2\hat{b}^2$, where the parameters $\chi_{a}$ and $\chi_{b}$ can be identified as third order susceptibilities of the optical nonlinear media -- we can refer to our oscillators as "Kerr-like" ones.  Such nonlinear systems can be a source of not only entangled states but also other relevant states commonly discussed in the papers devoted to the quantum optics problems. For instance, Miranowicz \textit{at al.} \cite{MTK90} have shown that Kerr-like oscillator can be a source of discrete superpositions of arbitrary number
of coherent states (so-called \textit{Schr\"odinger cat}-like or \textit{kitten} states). The oscillators  interact with each other by an nonlinear coupling of the strength $\epsilon$  -- the Hamiltonian corresponding to this interaction can be written as: $\hat{H}_i= \epsilon (\hat{a}^\dagger)^2\hat{b}^2+ \mathrm{h.c.}$. Additionally, one of the oscillators ($a$) is externally driven by a coherent field. The intensity of this linear interaction with the field is equal to $\alpha$ and the Hamiltonian corresponding to this process is $\hat{H}_e=\alpha \hat{a}^\dagger+ \mathrm{h.c.} $.

 While the system is not subjected to the interactions with any kind of external environment it can be treated as a {\em quantum scissors device} under some assumptions. The number of two-mode states engaged in the dynamics can be substantially truncated as both internal (characterised via $\epsilon$) and external (described by $\alpha$) interactions are small if compared  with nonlinearity parameters $\chi$. Therefore, as  explained in \cite{KL06}, the number of two-mode states effectively engaged in the system's dynamics is drastically reduced. The fidelity between such truncated state $|\Psi\rangle_{cut}=c_{02}|0\rangle_a|2\rangle_b+c_{12}|1\rangle_a|2\rangle_b+c_{20}|2\rangle_a|0\rangle_b$ and a two-mode state $|\Psi\rangle=\sum\limits_{n,m=0}^{\infty} c_{n,m}(t)\left|n\right>_a\left|m\right>_b$ is almost equal to  unity with the accuracy $\sim 10^{-4}$. We show this fact in Fig.1 where the fidelity is plotted for $\epsilon = 0.1$, $\alpha =\epsilon$ and $\chi = 25$, and the numerical calculations were performed in the $n$-photon Fock basis where $100$ two-mode states were involved.
 This result enables us to treat our system as a {\em nonlinear quantum scissors} device. In fact, such behaviour originates from the resonances between the energies of the levels generated by the Hamiltonian $\hat{H}_{nl}$ and the couplings discussed in the model. For further informations concerning linear and nonlinear quantum scissors devices see for example \cite{PPB98,LT94,KKGJ00,M05,LM04} \textit{and the references quoted therein} or the review papers \cite{MLI01,LM01}.

As  mentioned above, the system can beave as \textit{quantum scissors device} described in details in \cite{KL06}. This system is able to form Bell-like states that are examples of maximally entangled states (MES). It should be mentioned that for the cases when interactions with external field are absent, we deal practically with qubit-qubit system, whereas when this interaction is present our system should be treated as qutrit-qubit one. In consequence, thanks to the two interactions included in our model, we can generate the Bell-like states during the system evolution. They can be expressed in the following form:
\begin{eqnarray}
\left|B_1\right>&=&\frac{1}{\sqrt{2}}\left(\left|2\right>_a\left|0\right>_b
+i\,\left|0\right>_a\left|2\right>_b\right), \,\label{eq1}\\
\left|B_2\right>&=&\frac{1}{\sqrt{2}}\left(\left|2\right>_a\left|0\right>_b
-i\,\left|0\right>_a\left|2\right>_b\right),\, \label{eq2}\\
\left|B_3\right>&=&\frac{1}{\sqrt{2}}\left(\left|2\right>_a\left|0\right>_b
+\,\left|1\right>_a\left|2\right>_b\right)\,. \label{eq3}
\end{eqnarray}

Unfortunately, if we include the interaction of the system with external reservoir one can expect that such generation process becomes imperfect and we lose the entanglement in our system. 
It is so because the dynamics of the system cannot be restricted to the two-mode states which are engaged in $|\Psi\rangle_{cut}$ any more. The interactions with reservoir cause that the other two-mode states (the states with lower energy) can be populated and therefore, the system as a whole cannot be treated as a qutrit-qubit one. Moreover, due to the interactions with external field ($\alpha$) also the states with higher energy are populated, but to a lower extent, as this interaction is small (in comparison to $\chi$ value). The weakness of this interaction is a necessary condition for the application of the \textit{nonlinear scissors} mechanism.
In further considerations we shall concentrate on the initially populated subspace defined by the Bell-like state and  trace the entanglement within this subspace.
In such a way we are not interested in the entanglement defined for the whole system that can be produced during the interaction time, but in the previously occupied $3\otimes 2$ subsystem entanglement.

In further considerations we shall show that thanks to the proper choice of the external bath parameters we are able to influence both: degree and character of the entanglement decay, leading to various interesting phenomena.  Thus, in the present paper we shall concentrate on the squeezed vacuum state reservoir  and on the influence of the squeezing parameters on the entanglement decay processes.

To describe the evolution of the system influenced by a broadband squeezed vacuum we shall apply master equation approach for the reduced density matrix of the system. Including the standard \textit{rotating wave} and \textit{Markov} approximations we derive the master equation in the following form, analogously as in \cite{HO08}:
\begin{eqnarray}
\frac{d\hat{\rho}}{dt}&=&
{i}\left(\hat{\rho}\hat{H}-\hat{H}\hat{\rho}\right)+
\sum\limits_{j=a,b}\left[2\hat{C}_{j}\hat{\rho}\hat{C}^{\dagger}_{j}-\hat{C}^{\dagger}_{j}\hat{C}_{j}\hat{\rho}-\hat{\rho}\hat{C}^{\dagger}_{j}\hat{C}_{j}\right]\nonumber\\
&+&2\sum\limits_{j=a,b}\left[\hat{C}_{j}^{(n)\dagger}\hat{\rho}\hat{C}_{j}^{(n)}+\hat{C}_{j}^{(n)}\hat{\rho}\hat{C}_{j}^{(n)\dagger}-
\hat{C}_{j}^{(n)\dagger}\hat{C}_{j}^{(n)}\hat{\rho}-\hat{\rho}\hat{C}_{j}^{(n)}\hat{C}_{j}^{(n)\dagger}\right]\\
&-&\sum\limits_{j=a,b}\left[2\hat{C}_{j}^{(s)\dagger}\hat{\rho}\hat{C}_{j}^{(s)\dagger}-\hat{C}_{j}^{(s)\dagger}\hat{C}_{j}^{(s)\dagger}\hat{\rho}-
\hat{\rho}\hat{C}_{j}^{(s)\dagger}\hat{C}_{j}^{(s)\dagger}\right]
\nonumber\\
&-&\sum\limits_{j=a,b}\left[2\hat{C}_{j}^{(s)}\hat{\rho}\hat{C}_{j}^{(s)}-\hat{C}_{j}^{(s)}\hat{C}_{j}^{(s)}\hat{\rho}-
\hat{\rho}\hat{C}_{j}^{(s)}\hat{C}_{j}^{(s)}\right]
\nonumber
\,\,\, ,
\label{master}
\end{eqnarray}
where $\hat{H}=\hat{H}_{nl}+\hat{H}_i+\hat{H}_e$, the operators $\hat{C}_j$, $\hat{C}_j^{(s)}$ and $\hat{C}_j^{(n)}$   $(j=a,b)$ describe the damping in the modes $a$ and $b$, respectively, and are defined as: 
\begin{eqnarray}
\label{eq12}
\hat{C}_{j}=\sqrt{\gamma_{j}}\:\hat{j}\,\,\, ,\,\,\,\\
\hat{C}_{j}^{(n)}=\sqrt{\gamma_{j}N_j}\:\hat{j}\,\,\, ,\\
\hat{C}_{j}^{(s)}=\sqrt{\gamma_{j}M_j}\:\hat{j}\,\,\, ,\\
M_j=\sqrt{N_j(N_j+1)}\exp(-i\phi)
\end{eqnarray}
The parameters $\gamma_j$ are the spontaneous emission rates for both of the modes ($j=\{a,b\}$) , $M_j$ characterises degree of squeezing of external bath, whereas $\phi$ is the phase of squeezing for the bath field. If we assume that $M_j=0$ we deal with usual thermal bath characterized by the mean number of photons equal to $N_j$ in the modes $j=\{a,b\}$. However, if we look at the form of eq.(8) the strength of the squeezing can be characterised by the value of $N_j$ as well (and we shall use it in further considerations).

As  already pointed, the system populated under the interactions with external environment is more dimensional that considered at the beginning of the evolution ($3\otimes 2$ system). Anyway, we are interested in the dynamics within the subsystem which was occupied before the interactions were included. Therefore, to discuss the entanglement decay  we extract $3\otimes 2$ subspace occupied initially from the whole two-mode subspace which is involved in the system's dynamics,  using the truncation procedure. Of course, one should keep in mind that entanglement can be temporally transferred to other subspace, different from the subspace defined by the states (eqns.\ref{eq1}-\ref{eq3}).
The density matrix for the truncated subspace can be obtained from the full density matrix calculated for $6\times 6$. The truncation procedure is given by the following relation:
\begin{equation}
\rho_{trunc}=\left(|0\rangle\langle 0|+|1\rangle\langle 1|+|2\rangle\langle 2|\right)\rho\left(|0\rangle\langle 0|+|2\rangle\langle 2|\right)\,\,\, .
\end{equation}
In this way we trace the evolution of entanglement which, before the interactions with environment, lived exclusively in the $3\otimes 2$ space, keeping in mind that some amount of entanglement can be (and nodoubtly is) transferred periodically out of this traced subspace.

Since our system is not a simple qubit-qubit one (for such a model the \textit{concurrence} parameter for the entanglement description can be applied) but qutrit-qubit is considered, for determination of the degree of entanglement in the  $3\otimes 2$  subspace we use the \textit{negativity} parameter \cite{P96,HHH96,VW02}. It is defined using  the  eigenvalues ($\lambda_j$) of the partial transpose, with respect to the smaller-dimensional subsystem, of the truncated system's density matrix ($\rho^T_{trunc}$) as follows:
\begin{equation}
N(\rho)= max\left(0,-2\lambda_{min}\right)\,\, ,
\end{equation}
where $\lambda_{min}$ is the smallest of the eigenvalues of the partially transposed  density matrix of the system described.
Negativity was proved to be an entanglement monotone \cite{VW02} and as such it can be used as a good measure of entanglement. It takes a value of 1 for the MES states (for the case discussed here -- Bell-like states) and for separable ones it is equal to zero. Using such an entanglement measure we are able to determine the degree of entanglement in the whole system considered. 

As already pointed out in \cite{KL06,KL09}, the system under consideration can be a source of maximally entangled states $|B_1\rangle$, $|B_2\rangle$ and $|B_3\rangle$ (\textit{see} eqns.(\ref{eq1} -- \ref{eq3})). It should be mentioned that although in the paper \cite{KL09} the entanglement dynamics in thermal and zero-temperature reservoirs has already been presented, however for the entanglement within two-qubit subspace, in contrast to the situation dealt with here, where the $3\otimes 2$ system is considered. As shown in \cite{KL09}, the entanglement sudden death and rebirth phenomena within the $2\otimes 2$ subsystems considered were caused by the interaction of this subsystem with the third entangled state $|B_3\rangle$ from which the entanglement can be periodically transferred to the two qubit subsystem (formed by $|B_1\rangle$ and $|B_2\rangle$ states). We should point out that in the present paper we deal with entanglement generated in the qutrit-qubit ($3\otimes 2$) system and we shall concentrate on the influence of squeezed reservoir on the entanglement decay process.  

For the system considered here we assume that the MES (Bell-like state $|B_3\rangle$) is generated initially and next, the external squeezed bath starts to influence the system.
Since we are interested in the character of negativity decay when the system interacts with squeezed reservoir and especially, we want to check whether the squeezed reservoir can influence this decay in any way by examining the time-evolution of the negativity for various values of the parameters describing the reservoir squeezed field. As shown in our previous paper \cite{KL09}, if the Bell state interacts with zero-temperature reservoir and there is no interaction between the system's parts forming entangled state, the entanglement decays asymptotically i.e. for $t\rightarrow\infty$ $N(\rho)\rightarrow 0$. However, if we deal with the thermal reservoir, the sudden death of entanglement phenomenon occurs in the system (entanglement decays to zero in a finite time). 
For the system interacting with the broadband squeezed vacuum bath (the case discussed here) we see that the moment of time of the last sudden death of entanglement phenomenon occurrence can depend on the squeezing parameters.
In particular, for the system discussed here we assume that the bath is squeezed in the same mode as that in which the coupler is externally pumped. At first, looking at Fig.2 we can see the influence of the squeezing parameter $N_a$ on final disentanglement time $\tau_d$ for constant squeezed vacuum phase $\phi$. Similarly as for the case when we deal with not squeezed and zero-temperature bath ($N_a=0$) the time $\tau_d$ reaches high values -- we can conclude that we obtain almost an asymptotic decay of entanglement. However, higher values of $N_a$ make the sudden death of entanglement visible as finite values of time $\tau_d$ become considerably smaller. Shortening of $\tau_d$ can be observed for both cases considered here ($\phi = \{0, \pi\}$). Moreover, for these two cases the value of $\tau_d$ decreases continuously and tends asymptotically to its final value (\textit{see} Fig.2). However, comparing the two dependences of  $\tau_d$  on $N_a$, we can easily identify some values of $N_a$ for which $\tau_d$ is considerably different for $\phi = 0$ and $\phi=\pi$.

It means that by changing the value of $\phi$ we may prevent the system from rebuilding the entanglement despite the fact that there is still an interaction between both oscillators and $\alpha$ (strength of the interaction with  external pumping field) is large enough to produce another entanglement birth.

For further considerations we will choose non-zero values for $N_a$ and shall concentrate on the influence of the phase $\phi$ on the entanglement decay dynamics. In particular, we shall find such a value of $N_a$ for which we can observe the noticeable changes in the values of $\tau_d$ as $\phi$  varies form $0$ to $\pi$, concentrating in further considerations on the influence of the squeezed reservoir properties on rebuilding of the entanglement in the system.
For instance, if we neglect mutual interaction between two oscillators ($\epsilon =0$), assume weak external pumping ($\alpha =0.02)$ and $N_a=2$, $N_b=0$ (squeezed bath in one mode) sudden death of entanglement appears. 
We see  (Fig.3) that when the phase of squeezing is equal to zero (or $\pi$) the entanglement death appears slightly later than for the non-squeezed, thermal field case ($M_a=0$, $N_a=2$). However, if we additionally assume that apart from the absence of coupling between the oscillators the external coherent field is also cut off, the changes are practically unnoticeable and the result is practically identical to that of thermal reservoir case. Moreover, it is worth mentioning that any larger changes of the moment of time when the entanglement death appears are noticeable when the squeezing is present in the same mode of the reservoir modes that is externally pumped. 

At this point we assume that two oscillators interact with each other  (\textit{i.e.} $\epsilon\neq 0$) and we will focus on the differences in the entanglement decay due to the interactions with broadband squeezed vacuum, in comparison to the standard thermal reservoir. As shown in \cite{KL09} the oscillations of entanglement appear in the system and for the thermal reservoir case besides the regular oscillations (characteristic for zero-temperature environment) it is possible to observe sequences of the sudden deaths and entanglement revivals. But what is more interesting, for the case of the squeezed external bath the phase of the squeezed reservoir mode can influence the time after which the final disentanglement is obtained and also influences the amount of the entanglement rebuilt  in the system.
In particular, we have noticed that by changing the phase of the squeezed reservoir mode we can qualitatively change the decay of negativity.
It is seen for the whole range of $\epsilon/\alpha$ (for all cases: $\epsilon/\alpha>1$, $\epsilon/\alpha<1$ and $\epsilon/\alpha=1$). In particular, in Fig.4(a--c) we plotted  the time after which the final entanglement death occurs ($\tau_d$) as a function of squeezed bath phase $\phi$ showing that
one additional rebuilt maximum in negativity versus time appears for some range of squeezed reservoir phases $\langle \sim 0; \sim 12\pi/20)$ and $( \sim 31\pi/20;\sim 2\pi\rangle$ (a similar feature can be observed for the thermal reservoir case, although there \cite{KL09} the mean number of photons and external coupling strength dependences were discussed instead of phase dependence). For the system discussed here we can observe a significant shortening ($\sim 20\%$) of total disentanglement time $\tau_d$ for $\epsilon/\alpha=1$ (Fig.4b) and shortening by $\sim 17\%$ for both $\epsilon/\alpha>1$ and $\epsilon/\alpha<1$ (see Figs.4a and 4c).
In Fig.2 we can see that this shortening (when we compare the cases for $\phi=\pi$ and $\phi=0$) varies from $\sim 4\%$ to $\sim 24\%$ for various values of squeezing parameter $N_a$. Our considerations concern one of these $N_a$ values that cause the maximal $\tau_d$ shortening.

Moreover, it is seen that for all cases shown in Fig.4a--c for some values of $\phi$ we observe one entanglement rebirth event less as a result of the influence of the squeezing phase on the entanglement dynamics. This change is of sudden character and indicates that we have some sort of  the "phase transition" within the system discussed. 
For $\epsilon/\alpha>1$ and $\epsilon/\alpha<1$ (Fig.4a and Fig.4c) we can see that 
the interaction of the system with the squeezed vacuum (for $\phi\in (\sim 12\pi/20;\sim 31\pi/20)$) can shorten the time $\tau_d$ as we compare it with the case when ordinary thermal bath is considered. However this shortening is less pronounced -- see the inset. Beyond this range even though $\alpha$ is not strong enough to produce another entanglement rebirth, the interactions with the squeezed bath lead to the appearance of additional negativity re-occurrence and hence, to a significant ($\sim 17\%$) lengthen of $\tau_d$. When $\epsilon/\alpha=1$ (Fig.4b) the changes of $\phi$ are of the ''phase transition'' character again (as for the previous cases), but for this case one can observe vanishing of the last negativity maximum and consequently, a rapid shortening of $\tau_d$ time. 

The character of this ''transition'' is related to the fact that if the squeezing phase $\phi$ changes its value the terms in the master equation proportional to the $\exp(\pm i\phi)$ add or subtract each other. In consequence, the populations and coherences in the system considered (and corresponding to them matrix elements of $\rho$) will be affected as well.  Due to the fact that the definition of the \textit{negativity} involves \textit{max} function, changes in the matrix elements of $\rho$ lead to the rapid changes (zeroing) in the \textit{negativity}. A similar behaviour was discussed in \cite{JAB01} where the squeezing phase dependence of the decoherence rate was considered for the atomic system located inside the cavity.

 Thus, we can see that the phase of squeezed reservoir can influence not only the time of entanglement death but also can change the system's ability to rebuild the entanglement. Especially, we can see that even though the value of $\alpha$ is enough to cause the entanglement rebuild, the appropriately chosen phase $\phi$ may prevent the last entanglement reconstruction. 

One more aspect that can be influenced by the squeezed reservoir state phase $\phi$ is the amount of entanglement rebuild in the system during the last entanglement rebirth phenomenon observed. Fig.5 shows the dependence of the maximal value of the last $\mathcal{N}_{max_{last}}$ and last but one negativity maximum $\mathcal{N}_{max}$ on the phase $\phi$. Similarly, to the case discussed earlier (Fig.4) we can observe some kind of ''phase transition'' - like behaviour -- sudden changes of the maximal value of the last negativity reconstruction $\mathcal{N}_{max_{last}}$ for some values of $\phi$ (inset in Fig.5) when the amplitude of the last negativity maximum is equal to zero. One should note that for the same range of $\phi$ we observe significant shortening of the time $\tau_d$ as well. Additionally, we can see from Fig.5 that the interaction with the squeezed bath can reduce the amount of entanglement preserved in the system. This reduction depends on the value of $\epsilon/\alpha>1$ --   for $\epsilon/\alpha>1$ and $\epsilon/\alpha<1$ it is about $ 30\%$, while for $\epsilon/\alpha=1$ even $ 56\%$. 

For the case of smaller squeezing parameter ($N_a=1$, mode b is not squeezed) we can only observe the regular changes in the time after which entanglement disappears and we do not see any "phase transitions" in the system (Fig.6). The shortening of the death time is only by about $4\%$ and all the rebuild maxima of the negativity -- time dependence are present for all squeezed reservoir phases --- see Fig.6. It confirms the earlier conclusion  about the $\tau_d$ dependence on the squeezing parameter $N_a$ visible at Fig.2. It should be also mentioned that for this case the changes in value of the last rebuilt maximum are of the same character as for the case of stronger squeezing, but  smaller in  percent.

As we look at the time-evolution of both: negativity and populations of individual states, the question arises if  the reservoir effects and population transfer could influence considerably the negativity evolution. The answer can be obtained for instance, from  analysis of the trace of the density matrix $\rho_{trunc}$ corresponding to the $3\otimes 2$ subsystem considered. Thus, Fig.7 shows the plots of the negativities and the trace of $\rho_{trunc}$ for $\phi=0$ and $\pi$. We see that for both values of $\phi$ the whole system that was initially populated in this subspace ($Tr(\rho_{trunc})=1$ for t=0) starts to populate other states --- $Tr(\rho_{trunc})$ decreases and is practically identical for both values of $\phi$. The character of this decay depends on the value of the negativity for a given moment of time. In particular, the decay rate increases when the negativity goes to zero, whereas it  slows down as the negativity reaches its maximum (this effect is visible in particular for the first negativity reconstruction). However those effects do not play crucial role, especially for longer times. This  means that the population leakage can effect the entanglement dynamics to some extend.

\section{Summary}

We have analysed the influence of squeezed reservoir on the entanglement decay for a nonlinear system initially prepared in a maximally entangled state (one of the Bell-like state). Such a state determines some subspace of the states that can be  treated as qutrit-qubit system and we have concentrated on it. 
Obviously, when the interaction with an external reservoir is included, other states are populated and in that way the whole system is no more a simple qutrit-qubit one, but more dimensional qudits should been considered. Anyway, our aim was to trace the evolution of the entanglement within the initially occupied subspace. Therefore, we look for the entanglement evolution inside the same qutrit-qubit system, keeping in mind that the other states (corresponding to the lower energies of the system) are involved in the dynamics and some loses of the entanglement in the subspace considered are related to the populating of these states. 
For the case when the oscillators could not interact with each other we could observe only small variations in the values of time $\tau_d$ when the entanglement disappears completely as a function of the squeezing phase. Moreover, this value does not differs considerably from that for the model involving standard thermal reservoir. However, the more pronounced changes become visible  when both the interactions (external characterised by $\alpha$ and internal characterised by $\epsilon$) are present. From the discussion of a similar model \cite{KL09} (where the thermal bath was considered) we expect oscillations of negativity, but their character  depends on the squeezed reservoir properties. In particular, we have shown that  shortening of the death time  for the parameters used, by about $17\%$ to $20\%$ of its original value can be obtained while $\epsilon/\alpha$ changes for some range of squeezed reservoir phases. The changes in this shortening can be of the step character resembling in some sense a "phase transition" in the system. In fact, we observe one entanglement revival less or more, as a result of the interaction with the squeezed bath depending on the phase of squeezing. 
Moreover, some changes in the amplitude of the last negativity maximum due to the squeezing phase variations could be also observed. In consequence, we can reduce the value of the entanglement revival at the end of its life by about $30\%$ to $56\%$ of its original value. Concluding, we can say that the squeezing parameters of external bath can influence considerably the entanglement decay dynamics  leading to various interesting features.

\begin{figure}[h]
\begin{center}
\vspace*{-0.1cm}
\resizebox{15cm}{8cm}
                {\includegraphics{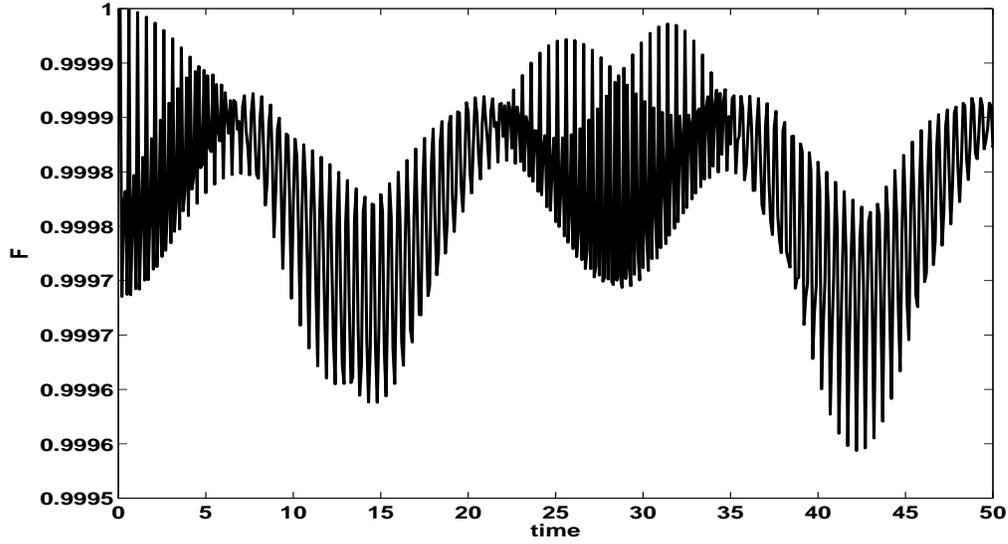}}
\vspace*{-0.5cm} 
\caption{The fidelity $F$ between the two-mode state $|\Psi\rangle=\sum\limits_{n,m=0}^{\infty} c_{n,m}(t)\left|n\right>_a\left|m\right>_b$ and a state corresponding to the qutrit-qubit system $|\Psi\rangle_{cut}=c_{02}|0\rangle_a|2\rangle_b+c_{12}|1\rangle_a|2\rangle_b+c_{20}|2\rangle_a|0\rangle_b$. The parameters: $\chi_a=\chi_b=25$, $\alpha/\epsilon=1$, $\epsilon=0.1$, $\gamma_a=\gamma_b=0$; time is scaled in $1/\chi$ units.} 
\end{center}
\end{figure}

\begin{figure}[h]
\begin{center}
\vspace*{-0.1cm}
\resizebox{15cm}{8cm}
                {\includegraphics{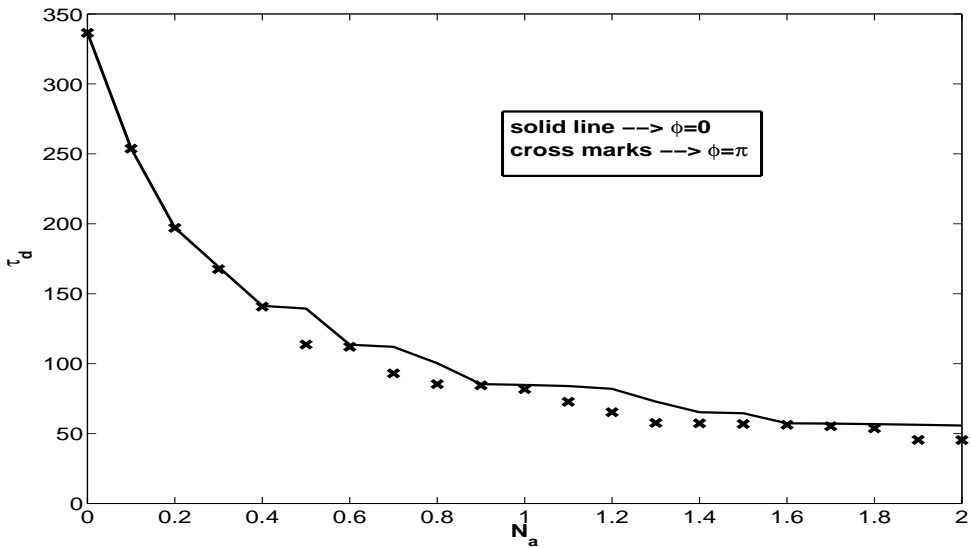}}
\vspace*{-0.5cm} 
\caption{Time of total entanglement death $\tau_d$ (scaled in $1/\chi$ units) versus the squeezing strength characterised by $N_a$. $\chi_a=\chi_b=25$, $\alpha=0.01$, $\epsilon=0.1$, $\gamma_a=\gamma_b=0.0025$. We assume that $|\Psi(t=0)\rangle=|B_3\rangle$}
\end{center}
\end{figure}

\begin{figure}[h]
\begin{center}
\vspace*{-0.1cm}
\resizebox{15cm}{8cm}
                {\includegraphics{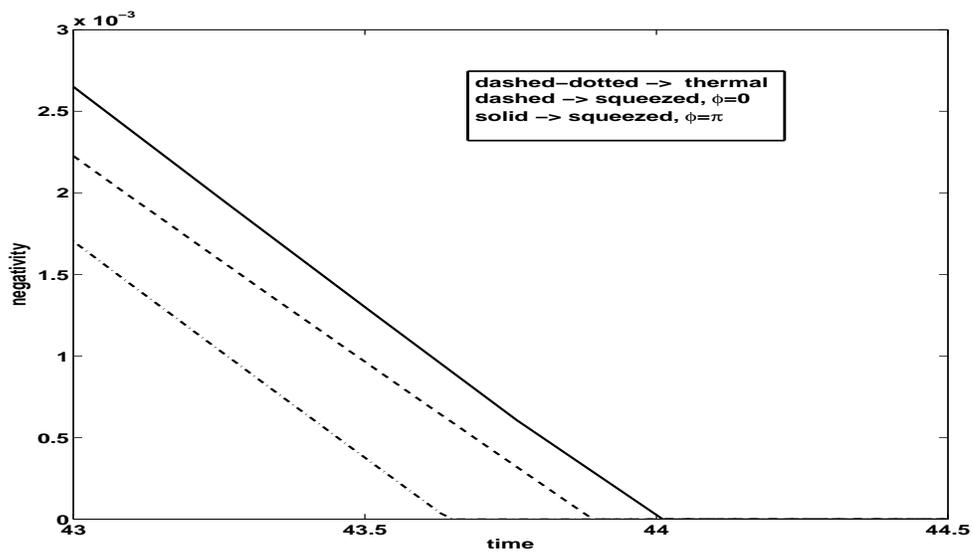}}
\vspace*{-0.5cm} 
\caption{Negativity $N(\rho)$ versus time scaled in $1/\chi$ units; $\chi_a=\chi_b=25$, $\alpha=0.01$, $\epsilon=0$, $\gamma_a=\gamma_b=0.0025$ and $N_a=2$, $N_b=0$, $|\Psi\rangle_in=|B_3\rangle$.}
\end{center}
\end{figure}

\begin{figure}
\resizebox{8cm}{8cm}
                {\includegraphics{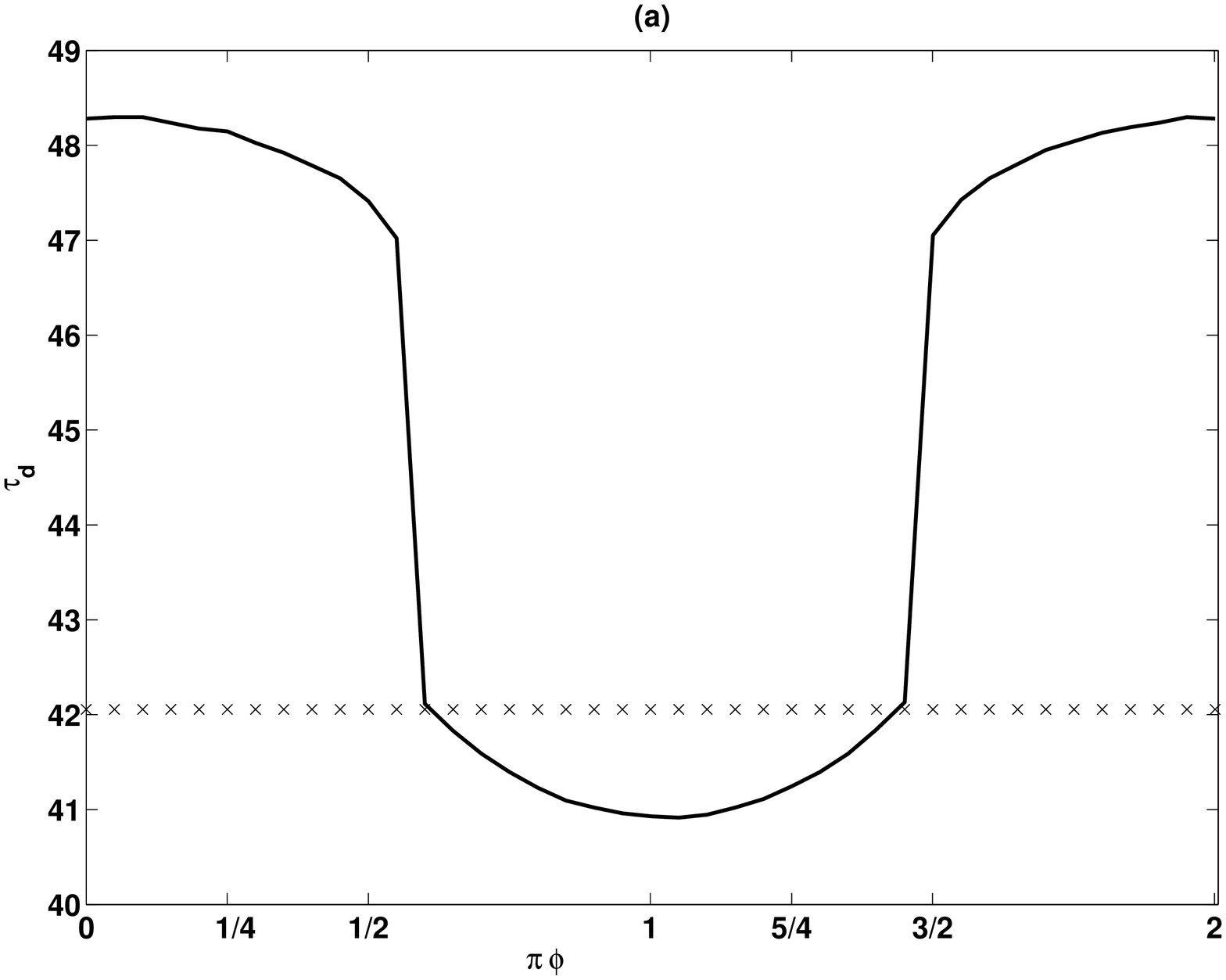}}
\resizebox{8cm}{8cm}
                {\includegraphics{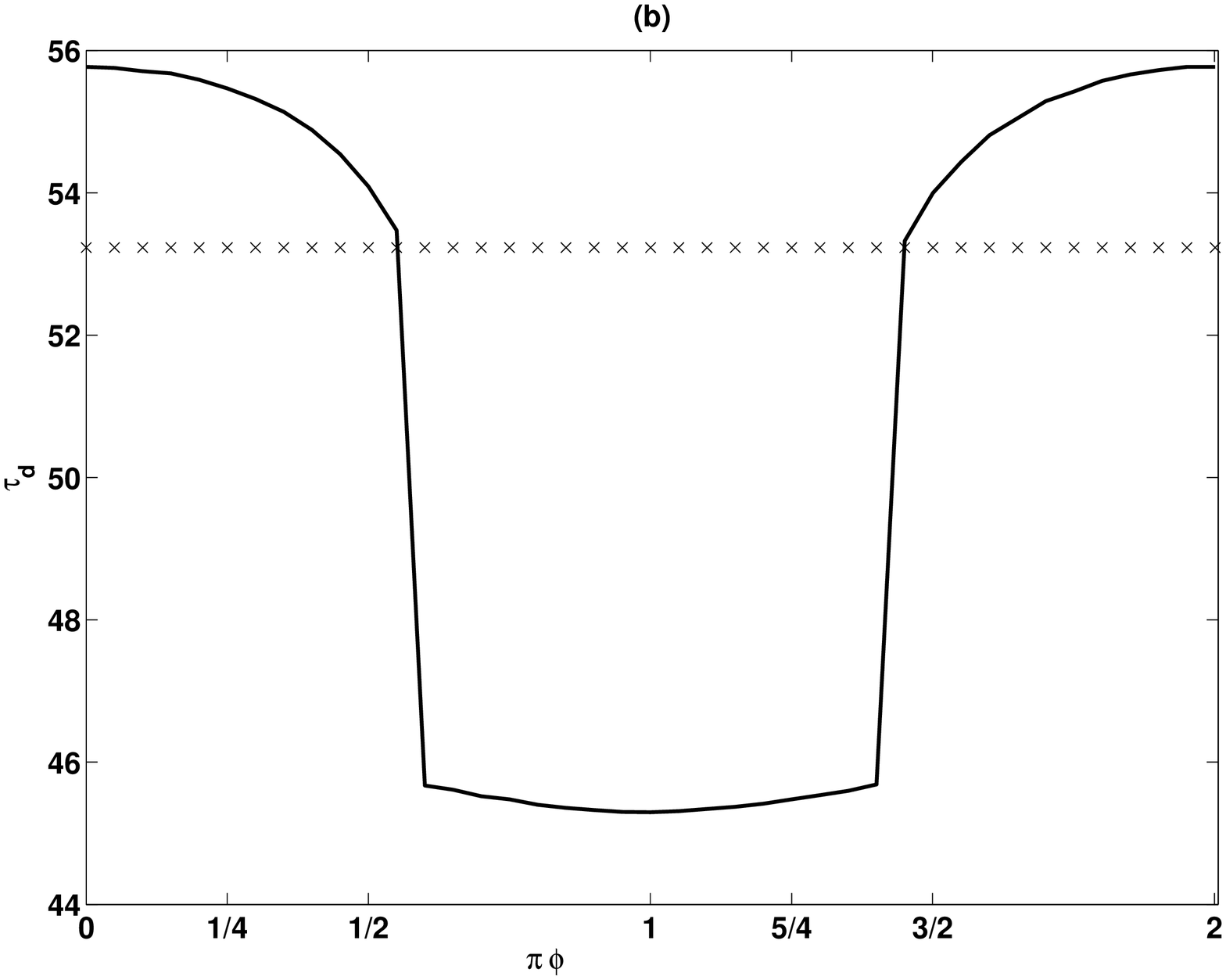}}
\resizebox{8cm}{8cm}
                {\includegraphics{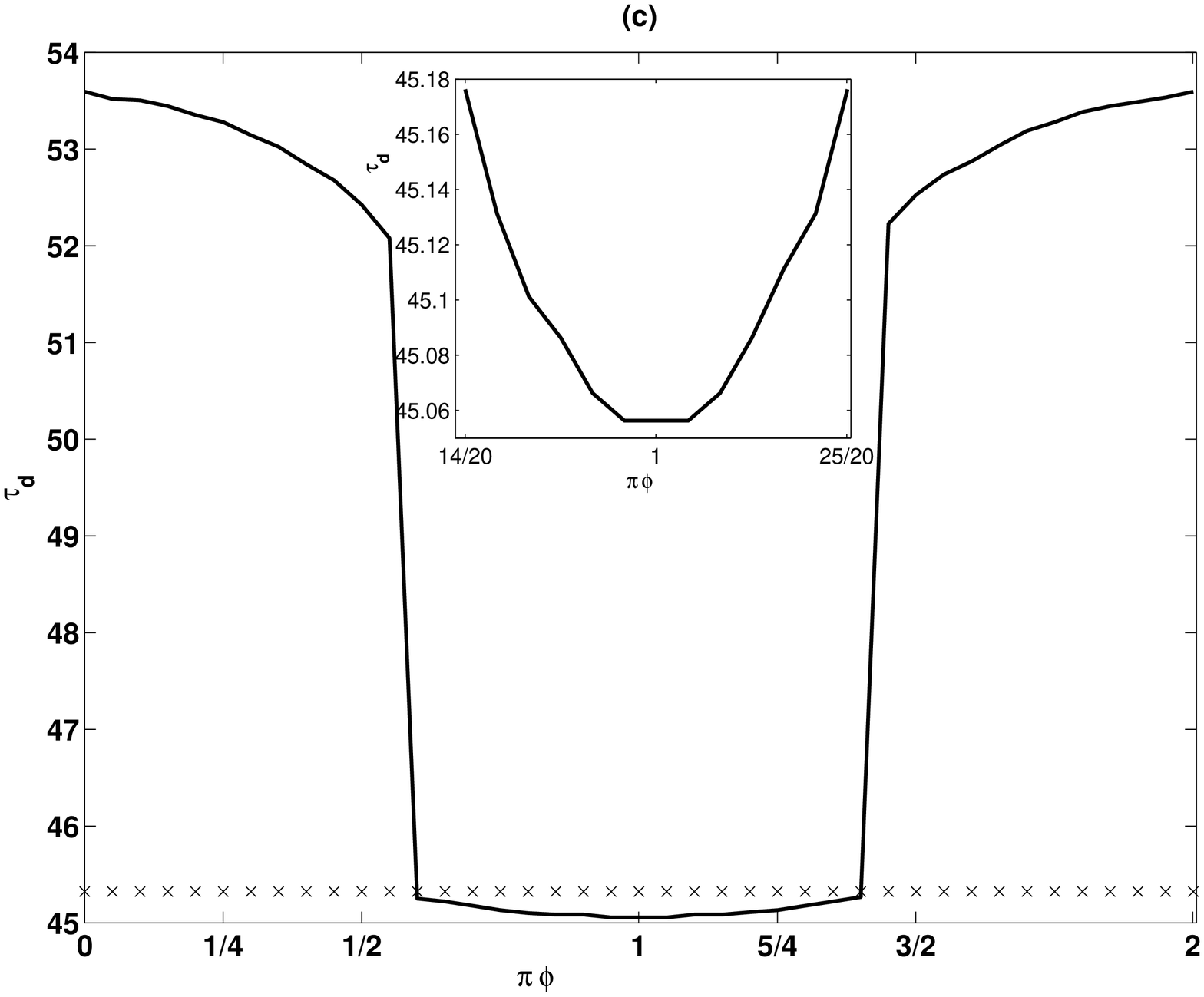}}
\caption{The dependence of $\tau_d$ as a function of the squeezing  phase $\phi$. The system is initially prepared in a $|B_3\rangle$ state. The time is scaled in $1/\chi$ units, $\chi_a=\chi_b=25$, $\epsilon=0.1$ , $\gamma_a=\gamma_b=0.0025$ and $N_a=2$, $N_b=0$.
At (a) $\epsilon/\alpha=10$, (b) $\rightarrow \epsilon/\alpha=1$ and at (c) $\epsilon/\alpha=0.5$. 
 Inset in (c) shows the central part of the whole plot. Stars correspond to thermal non-squeezed reservoir characterized by the same mean number of photons.}
\end{figure}

\begin{figure}[h]
\begin{center}
\vspace*{-0.1cm}
\resizebox{14cm}{8cm}
                {\includegraphics{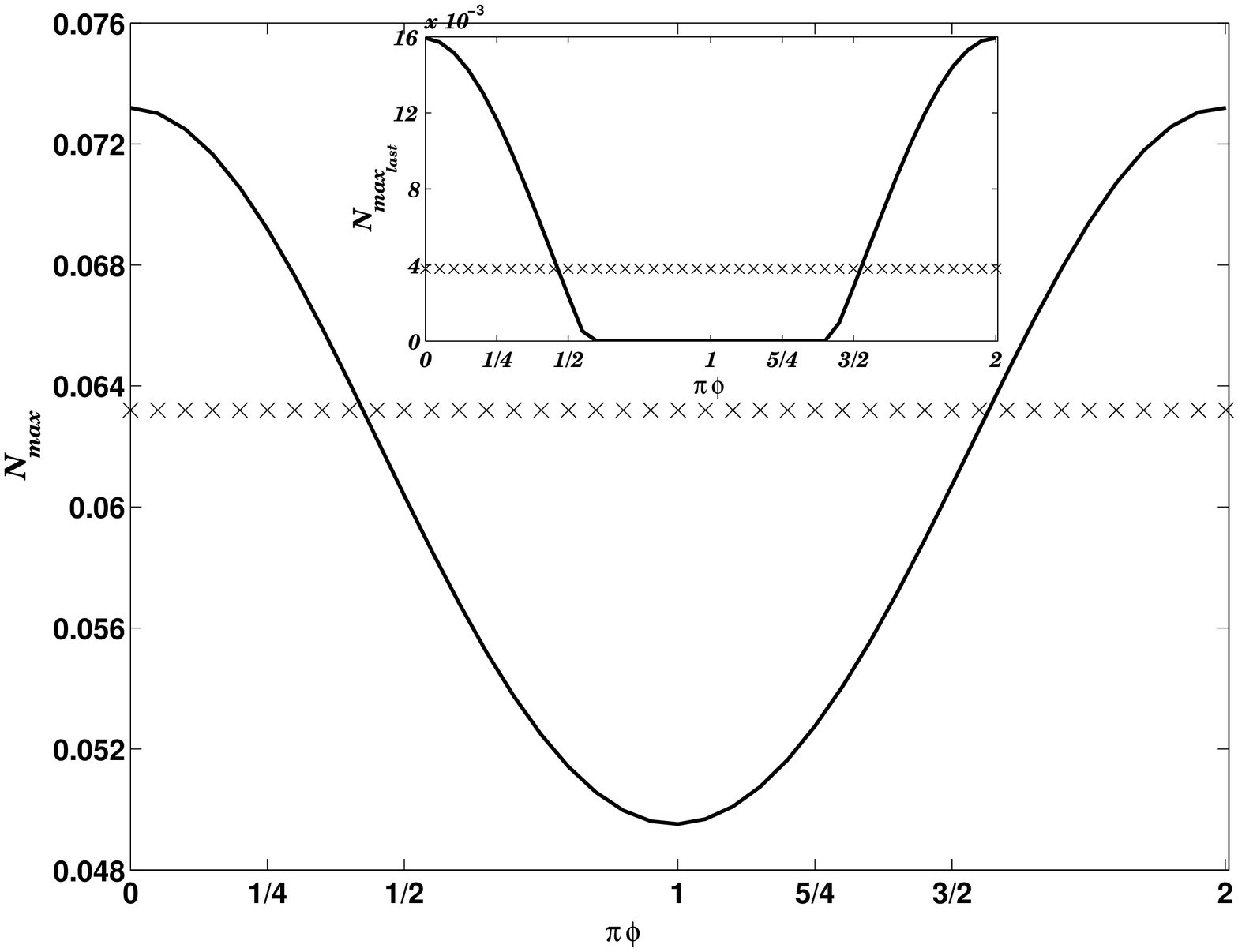}}
\vspace*{-0.5cm} 
\caption{The maximal value of the  last but one negativity reconstruction $\mathcal{N}_{max}$ for the entanglement reborn appearance as a function of phase of squeezing reservoir $\phi$. The system parameters are the same as in Fig.4b. At the inset the amplitude of the last negativity maximum  $\mathcal{N}_{max_{last}}$.}
\end{center}
\end{figure}

\begin{figure}[h]
\begin{center}
\vspace*{-0.1cm}
\resizebox{15cm}{10cm}
                {\includegraphics{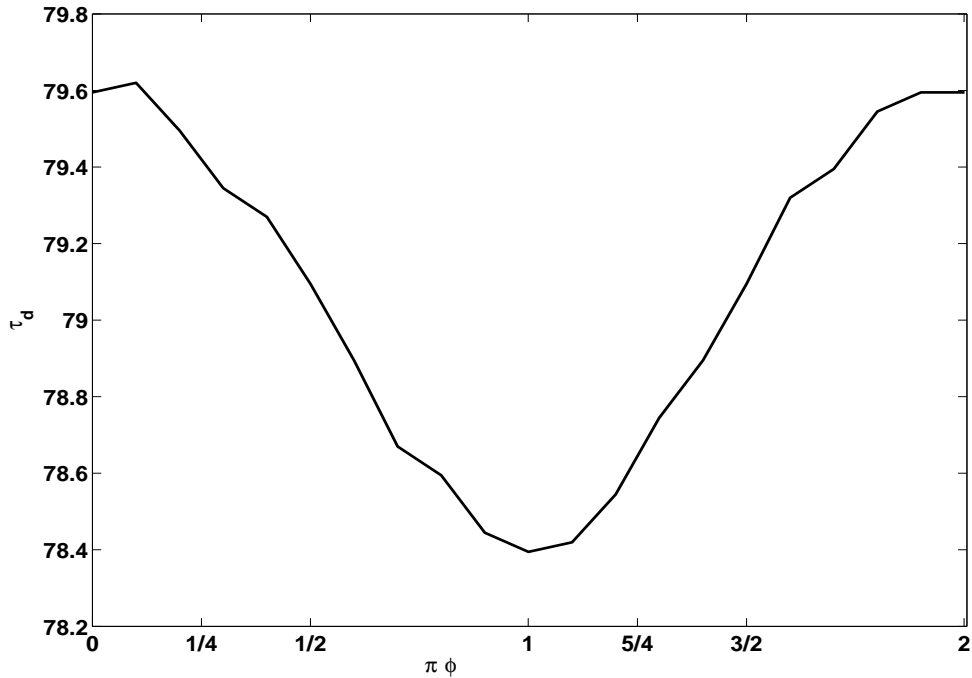}}
\vspace*{-0.5cm} 
\caption{The same as in Fig.4c but for $N_a=1$ and $N_b=0$.}
\end{center}
\end{figure}

\begin{figure}[h]
\begin{center}
\vspace*{-0.1cm}
\resizebox{15cm}{10cm}
                {\includegraphics{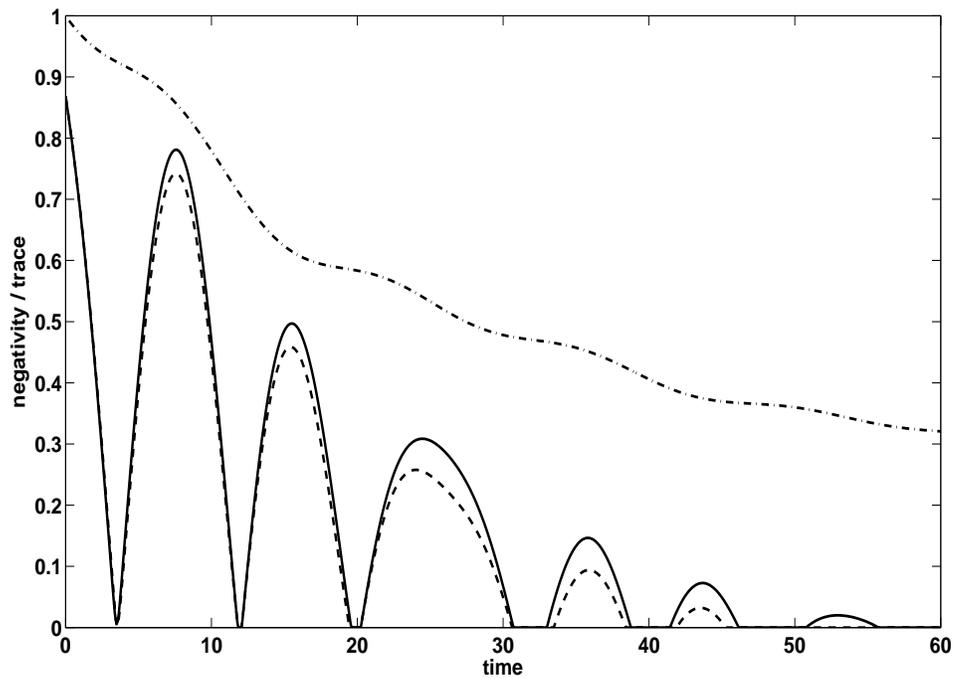}}
\vspace*{-0.5cm} 
\caption{Negativity for the qutrit-qubit subsystem described by the density matrix $\rho_{trunc}$ (eq.(9)) for $\phi=0$ (solid line), $\phi=\pi$ (dashed line) and the trace of $\rho_{trunc}$ (dashed-dotted line). The remaining parameters are the same  as in Fig.4b. The trace of $\rho_{trunc}$ is identical for both values of $\phi$.}
\end{center}
\end{figure}

\end{document}